# VIES MOYENNES DES ETATS EXCITES DU $^{19}$F

A. BELAFRITES[1], A. BOUCENNA[2], C. BENAZZOUZ[3] et M. A. BENAMMAR[3]

[1]Département de physique, Centre Universitaire de Jijel, BP 98 Ouled Aissa JIJEL 18000 Algérie.
[2]Département de physique, Faculté des Sciences Université Ferhat Abbas Sétif 19000 Algérie.
[3]Centre de Développement des Techniques Nucléaires C.D.TN. 16000 Alger Algérie.

**RESUME**

La détermination des vies moyennes des états excités a été entreprise par l'excitation coulombienne, qui est un processus de diffusion inélastique dans lequel la particule incidente est chargée et provoque une excitation du noyau cible par l'intermédiaire du champs électromagnétique durant le passage du projectile. La section efficace de cette diffusion inélastique est proportionnelle à la largeur de l'état isomérique. Les vies moyennes du $^{19}$F sont déterminées en comparant les résultats expérimentaux des réactions $^{19}$F($\alpha,\alpha'\gamma$)$^{19}$F et $^{19}$F(p,p'$\gamma$)$^{19}$F aux calculs théoriques réalisées à l'aide du programme de l'excitation coulombienne multiple Coulex. Les vies moyennes sont comparées à celles retenues par Ajzenberg –Selove [1] et Alder et al. [2].

The determination of the mean lifetime of excited states has been undertaken by coulomb excitation that is a process of inelastic scattering in which the incident particle is charged and provokes an excitation of the target nucleus by mean of the electromagnetic fields during the passage of the projectile. The cross section of this inelastic scattering is proportional to the width of the isomeric state. The mean lifetime of excited states of the $^{19}$F are determined by comparing the experimental results of reactions $^{19}$F($\alpha,\alpha`\gamma$)$^{19}$F and $^{19}$F(p, p`$\gamma$)$^{19}$F with the theoretical calculations achieved with the help of the multiple coulomb excitation program Coulex. These mean lifetimes are compared to those restraints by Ajzenberg - Selove [1] and Alder and al. [2].

## I. INTRODUCTION

L'excitation coulombienne des noyaux atomiques est un moyen adapté pour déterminer les valeurs des éléments de matrice réduits des noyaux. La section efficace de cette diffusion inélastique est au premier ordre de perturbation, proportionnelle à la probabilité de désexcitation de l'état excité. Sa mesure permet de déduire la vie moyenne de l'état excité [3]. Dans ce travail, nous exploitons la technique de l'excitation coulombienne du $^{19}$F par les protons et les noyaux d'Hélium (Alpha) pour déterminer les vies moyennes des états excités du noyau cible $^{19}$F. Les calculs Théoriques ont été réalisés à l'aide du programme Coulex [4], [5] et [6]. Les expériences ont été effectuées au niveau du Centre de Développement des Techniques Nucléaires C.D.T.N d'Alger.

## II. DONNEES EXPERIMENTALES

### 1. Réaction $^{19}$F($\alpha, \alpha`\gamma$)$^{19}$F

L'excitation coulombienne du $^{19}$F est réalisée en bombardant une cible mince de CaF$_2$ par des particules alpha d'énergie $E_\alpha$ = 2.5 MeV, et de 400 nA d'intensité délivrée par l'accélérateur Van de Graaff 3.75 MeV du C.D.T.N d'Alger. Les rayons gamma émis de cette réaction sont détectés par un détecteur germanium hyper pur. Le spectre obtenu est montré sur la figure 1.

Les résultats de dépouillement de ce spectre sont reportés dans le tableau 1. Sur ce spectre, on distingue deux pics correspondant aux transitions gamma E1 d'énergie 110 keV et E2 d'énergie 197 keV, et un pic à 87 keV correspondant à la transition E3.

| N° Pic | Energie (keV) | Aire | FWHM |
|---|---|---|---|
| 1 | 197 | 3636 | 3.23 |
| 2 | 110 | 67 | 4.06 |
| 3 | 87 | 214 | 14.86 |

**Tableau 1**
**Donnée expérimentales déduites du spectre**
**de la réaction $^{19}F(\alpha, \alpha'\gamma)^{19}F$ à l'énergie $E_\alpha = 2.5$ MeV**

Par ailleurs Temmer et al[7] ont bombardé une cible de $CaF_2$ par des particules alpha d'énergie incidente $E_\alpha = 1.8$ MeV et ont détecté les rayons gamma émis par un détecteur à scintillations NaI(Tl) (2in x 2in). Le spectre obtenu est donné dans la figure 2. Sur ce spectre, on distingue deux pics correspondant aux transitions gamma E1 d'énergie 113 keV et E2 d'énergie 196 keV et on remarque l'absence du troisième pic d'énergie à 83 keV, et par conséquent l'absence de la transition E3. Les résultats extraits du spectre de la réaction $^{19}F(\alpha,\alpha`\gamma)^{19}F$ sont reportés au tableau 2. La figure 3 montre les rendements relatifs des rayons gamma en fonction de l'énergie incidente des particules alpha $E_\alpha$.

| N° Pic | Energie (keV) | Aire | Rendement( $E_\alpha = 1.8$ MeV ) |
|---|---|---|---|
| 1 | 196 | 520 | 74 |
| 2 | 113 | 41 | 4 |
| 3 | 83 | - | - |

**Tableau 2**
**Donnée expérimentales déduites du spectre**
**de la réaction $^{19}F(\alpha, \alpha'\gamma)^{19}F$ à $E_\alpha$=1.8 MeV**

Les sections efficaces expérimentales de la réaction $^{19}F(\alpha, \alpha'\gamma)^{19}F$ obtenues par Sherr et al. [8] et Alder et al. [2] obtenus aux énergies incidentes $0.8 \leq E_\alpha \leq 20$ MeV sont montrées sur la figure 4.

## 1. Réaction $^{19}F(p, p'\gamma)^{19}F$

Une cible épaisse (pastille) de 0.5 mm d'épaisseur de LiF a été bombardée par un faisceau de protons de 1.6 MeV d'énergie et de 30 nA d'intensité délivrée par l'accélérateur Van de Graaff 3.75 MeV du C.D.T.N d'Alger. Les rayons gamma émis par cette réaction sont détectés par un détecteur germanium de haute pureté GeHP placé à une distance de 30 cm de la cible, formant un angle $\Theta = 165°$ par rapport à la direction du faisceau incident. L'étalonnage de la chaîne de spectroscopie gamma est faite par deux sources d'énergie connues, qui sont le $^{137}Cs$ ayant un pic à 662 keV et $^{241}Am$ ayant un pic à 60.26 keV. Sur le spectre (figure 5), on distingue trois pics correspondants aux transitions E1, E2 et E3, montrées sur le schéma de niveaux du $^{19}F$ (figure 6). La présence du pic à 87 KeV est en accord avec les données de N. P. Heydenberg et al. [9] obtenues par la réaction $^{19}$F($\alpha$, $\alpha'\gamma$)$^{19}$F. Ce pic à 87 KeV qui apparaît à 83 keV dans les résultats de N. P. Heydenberg [9] semble disparaître en utilisant une meilleure détection dans les données de Temmer et al.[7]. Ce pic à 83 keV a d'ailleurs troublé Peterson et al. [10] et Jones et al. [11]. La présence d'un pic à 87 KeV dans le spectre de la réaction $^{19}$F(p, p'$\gamma$)$^{19}$F que nous avons mesuré est certaine (figure 5).

Le dépouillement des données expérimentales a été réalisé à l'aide du logiciel spectra I [12]. L'analyse du spectre de la réaction $^{19}$F(p, p'$\gamma$)$^{19}$F donne les valeurs reportées dans le tableau 3.

| N° Pic | Energie (keV) | Aire | FWHM |
|---|---|---|---|
| 1 | 197 | 584 | 12.45 |
| 2 | 110 | 49 | 09.39 |
| 3 | 87 | 32 | 16.45 |

**Tableau 3**
**Donnée expérimentales déduites du spectre**
**de la réaction $^{19}$F(p, p`$\gamma$)$^{19}$F**

### III. CALCULS THEORIQUES

#### 1. Réaction $^{19}$F($\alpha$, $\alpha'\gamma$)$^{19}$F

Les calculs théoriques réalisés par Alder et al. [13] et Mulin et al [14] donnés dans l'étude de Temmer et al [7] sont représentés sur la figure 3 (courbe pleine). Ces calculs sont normalisés par rapport aux valeurs expérimentales correspondant à l'énergie incidente $E_\alpha = 1.0$ MeV.

Les valeurs théoriques des sections efficaces d'excitation de la réaction $^{19}$F($\alpha$,$\alpha'\gamma$)$^{19}$F calculées par Sherr et al. [8] sont montrées sur la figure 5 (courbes pleines). Elles sont normalisées à la valeur de la section efficace expérimentale correspondant à $E_\alpha = 1.55$ MeV.

A l'aide du programme de l'excitation coulombienne multiple Coulex et en utilisant les éléments de matrice réduits cités dans la littérature [2], [9] et [15] : $M(E1) = 0.0015e\sqrt{barn}$, $M(E2) = 0.05 ebarn$ et $M(E3) = 0.03e(barn)^{3/2}$, nous avons calculé les probabilités d'excitation des niveaux excités à 110 keV et à 197 keV du noyau $^{19}$F. Nous avons fait varier l'énergie $E_\alpha$ de 0.5 MeV jusqu'à 3.0 MeV par pas de 0.1 MeV avec l'angle $\Theta_\gamma = 165°$ et $\Theta_p = 160°$. Un autre calcul a été fait en variant l'angle $\Theta_p$ de 20° à 180° par un pas de 20° et en gardant $E_\alpha = 1.6$ MeV et $\Theta_\gamma = 165°$. Les probabilités d'excitation calculées sont données dans la figure 7. Dans la figure 8 sont montrées les courbes des sections efficaces théoriques normalisées séparément pour chaque niveau à $E_\alpha = 1.55$ MeV et comparées à celles obtenues dans l'étude de Sherr et al. [8].

#### 2. Réaction $^{19}$F(p, p'$\gamma$)$^{19}$F

Les calculs théoriques sont réalisés à l'aide du programme Coulex, en utilisant les mêmes éléments de matrice réduits. Pour des énergies variant de 0.3 MeV jusqu'à 2.0 MeV et les angles de la diffusion variant de 10° à 180°. Ces données conduisent au schéma de décroissance du noyau cible $^{19}$F (figure 5) et les rapports d'embranchement donnés dans le tableau 5, avec les vies moyennes $\tau_1$ = 1.4 nsec pour le niveau ½⁻ à 110 keV et $\tau_2$ = 220.1 nsec pour le niveau 5/2⁺ à 197 keV.

| Niveau (N) | Niveau (Q) | Transition | Energie (MeV) | branching |
|---|---|---|---|---|
| 2 | 1 | E1 | 0.110 | 0.10000E+01 |
| 3 | 1 | E2 | 0.197 | 0.10000E+01 |
| 3 | 2 | E3 | 0.087 | 0.21320E-08 |

**Tableau 4**
**Schéma de décroissance du noyau cible ¹⁹F et les rapports d'embranchement.**

Les probabilités d'excitation des niveaux du ¹⁹F en fonction de l'énergie des protons $E_P$ sont données dans la figure 9.

## IV. DETERMINATION DES VIES MOYENNES

Afin de comparer nos calculs avec les données expérimentales, nous avons considéré les rendements σ des réactions ¹⁹F(α, α'γ)¹⁹F et ¹⁹F(p, p'γ)¹⁹F et déduit les rapports des rendements reportés dans le tableau 5.

| Réaction | $\sigma_1$ (unités arbitraires) | | $\sigma_2$ (unités arbitraires) | | $\sigma_1/\sigma_2$ | |
|---|---|---|---|---|---|---|
| | expérience | Théorique | expérience | Théorique | expérience | Théorique |
| ¹⁹F(α,α'γ)¹⁹F | 520[1)] | 35 [1)] | 41[1)] | 18[2)] | 12.68[1)] | 14[1)] |
| | 52[1°)] | 380[2)] | 3.2[1°)] | 18[2)] | 16.2[1°)] | 21[2)] |
| | 380[2)] | 54[3)] | 18[2)] | 0.1[3)] | 21[2)] | 50.6[3)] |
| ¹⁹F(p, p'γ)¹⁹F | 584[3)] | 4.309[3)] | 49[2)] | 0.01895[3] | 11.9[3)] | 227.4[3)] |

1. Données de la référence [7], en intégrant le spectre de la ¹⁹F(α, α`γ)¹⁹F.
1°. Données de la référence [7].
2. Données des références [8] et [9].
3. Ce travail.

**Tableau 5**
**Rapports des rendements expérimentaux et théoriques entre les niveaux excités de spin 5/2⁺ et ½⁻ du ¹⁹F.**

Dans ce tableau, $\sigma_1$ représente les rendements du niveau 5/2⁺ à 197 keV et $\sigma_2$ représente les rendements du niveau ½⁻ à 110 keV, pour les énergies incidentes $E_\alpha$=1.6 MeV et $E_P$ =1.6 MeV. On constate que les valeurs expérimentales du rapport $\sigma_1/\sigma_2$ sont comprises entre 10 et 20. Les valeurs théoriques tirées des références [7], [8] et [9] ont été normalisées par rapport aux valeurs expérimentales. Les valeurs du rapport $\sigma_1/\sigma_2$ que nous avons calculées pour les réactions ¹⁹F(α, α'γ)¹⁹F et ¹⁹F(p, p'γ)¹⁹F présentent un désaccord considérable avec les valeurs expérimentales alors que l'accord est satisfaisant entre les valeurs relatives expérimentales et théoriques des sections efficaces (figure 8). Nous avons attribué ce désaccord entre les rapports expérimentaux et théoriques aux valeurs des éléments de matrices réduits et donc aux vies moyennes partielles que nous avons introduites initialement pour faire les calculs. L'ajustement des valeurs théoriques aux valeurs expérimentales nous permettra d'avoir une estimation des éléments de matrice réduits.

## 1. Vies moyennes

En utilisant les rendements $\sigma_1$ du niveau $5/2^+$ à 197 keV et $\sigma_2$ du niveau $½^-$ à 110 keV du $^{19}F$ déduits du spectre de la réaction $^{19}F(p, p'\gamma)^{19}F$, nous avons calculé des rapports expérimentaux $\sigma_1/\sigma_2$. A l'aide du programme Coulex, nous avons obtenu des sections efficaces pour ces mêmes niveaux et nous avons calculé ces rapports théoriques $\sigma_1/\sigma_2$. L'ajustement des valeurs théoriques de ces rapports aux valeurs expérimentales en faisant varier les éléments de matrice réduits et donc les vies moyennes partielles des niveaux, nous ont permis d'avoir une estimation des vies moyennes. Plusieurs jeux de paramètres M(E1) et M(E2) donnent un rapport théorique $\sigma_1/\sigma_2$ compris entre 10 et 20. Donc, plusieurs valeurs des éléments de matrice M(E1) et M(E2) donnent un rapport proche de l'expérience. En utilisant ces valeurs pour calculer les vies moyennes correspondantes, nous obtenons des valeurs comprises entre 73 nsec et 1834 nsec pour la vie moyenne du niveau à 197 keV et de 0.26 nsec à 3.15 nsec pour la vie moyenne du niveau à 110 keV.

Les valeurs des vies moyennes $\tau_1$ et $\tau_2$ qui s'accordent avec les vies moyennes retenues par Ajzenberg-Selove[1] sont : $\tau_1$ = 0.87 *nsec* et $\tau_2$ =136 *nsec* correspondent aux éléments de matrice réduits M(E1) = 0.0033 $e\sqrt{barn}$ et M(E2) = 0.1100 *ebarn* L'élément M(E3) correspondant à la valeur de B(E3) = 0.001 *ebarn*$^{3/2}$ donne une intensité meilleure pour la transition E3. Les vies moyennes déduites de la comparaison des rapports $\sigma_1/\sigma_2$ expérimentaux et théoriques montrent que le niveau $5/2^+$ se désexcite vers le niveau $1/2^-$ et vers le niveau fondamental $1/2^+$ et sa vie moyenne est déduite des vies moyennes correspondant aux transitions E2 et E3. Dans le tableau 6, on donne les valeurs des vies moyennes comparées à celles tirées des références [15], [2] et [1].

| Vie moyenne | Nos Résultats | [15] | [2] | [1] |
|---|---|---|---|---|
| $\tau_1$ | 0.8689 (nsec) | 0.89 (nsec) | 1.10 (nsec) | 0.87 (nsec) |
| $\tau_2$ | 136.56 (nsec) | 129 (nec) | 86.6 (nsec) | 128.9 (nsec) |

**Tableau 6**
**Vies moyennes des niveaux $5/2^+$ et $½^-$ du $^{19}F$**

## 2. Sections efficaces

Nous avons utilisé les valeurs des éléments de matrice réduits que nous avons déterminées pour calculer les sections efficaces théoriques de la réaction $^{19}F(\alpha, \alpha'\gamma)^{19}F$ pour laquelle, nous disposons des données expérimentales [8]. Le calcul est réalisé toujours à l'aide du programme Coulex et les courbes sont représentées dans la figure 10. Un seul point de normalisation des valeurs théoriques permet d'avoir un accord satisfaisant entre l'expérience et la théorie (figure 10) pour les deux niveaux à 110 keV et 197 keV alors qu'avec les anciennes valeurs des éléments de matrices réduits, il fallait normaliser séparément les sections efficaces correspondant à chacun des deux niveaux.

## V. CONCLUSION

L'excitation coulombienne du noyau $^{19}F$ a été entreprise en utilisant des faisceaux de protons et des particules alpha délivrés par l'accélérateur Van de Graaff 3.75 MeV. Le $^{19}F$ est un noyau légèrement déformé ayant un niveau $½^-$ à 110 keV et un niveau $5/2^+$ à 197 keV. Ces deux niveaux peuvent s'exciter en utilisant des protons d'énergie inférieure à l'énergie de sécurité maximum qui est de l'ordre de 1.6 MeV et des particules alpha d'énergie inférieure à 3.00 MeV, ce qui garantit l'origine électromagnétique de l'interaction mise en jeu. Les rendements des réactions $^{19}$F(p, p'γ)$^{19}$F et $^{19}$F(α, α'γ)$^{19}$F ont été mesurés et les probabilités d'excitation ont été calculées en utilisant le programme Coulex. L'ajustement des valeurs théoriques aux données expérimentales a permis de déterminer les vies moyennes $τ_1$ = 0.86 *nsec* et $τ_2$ = 136 *nsec* des niveaux à 110 keV et à 197 keV du $^{19}$F respectivement. Ces vies moyennes ont été déduites des éléments de matrice multipolaires du noyau cible $^{19}$F $M(E1) = 0.0033\ e\sqrt{barn}$, $M(E2) = 0.11\ ebarn$ et $M(E3) = 0.0445\ e(barn)^{3/2}$ qui ont donné le meilleur accord avec les rapports $σ_1/σ_2$ expérimentaux.

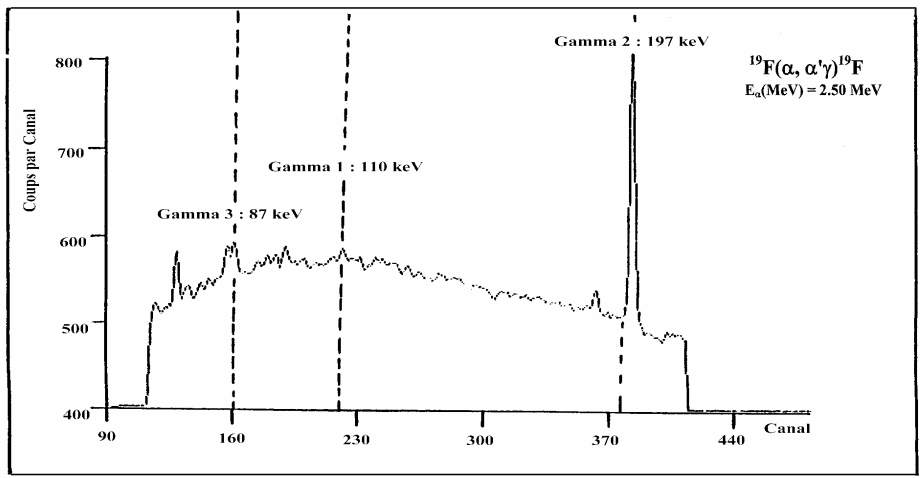

Fig.1: Spectre des rayons gamma émis de l'excitation coulombienne du $^{19}$F par des particules alpha d'énergie $E_\alpha$ = 2.50 MeV

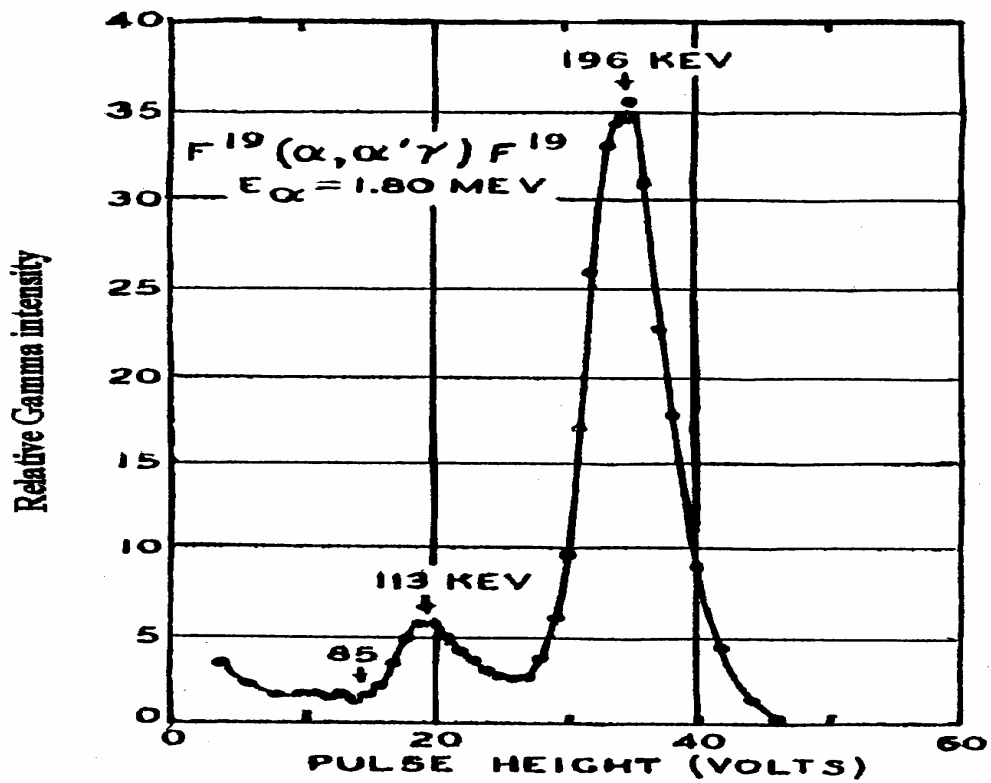

Fig. 2: Spectre des rayons gamma [7] émis de l'excitation coulombienne du $^{19}$F par les particules alpha d'énergie $E_\alpha$ = 1.8 MeV

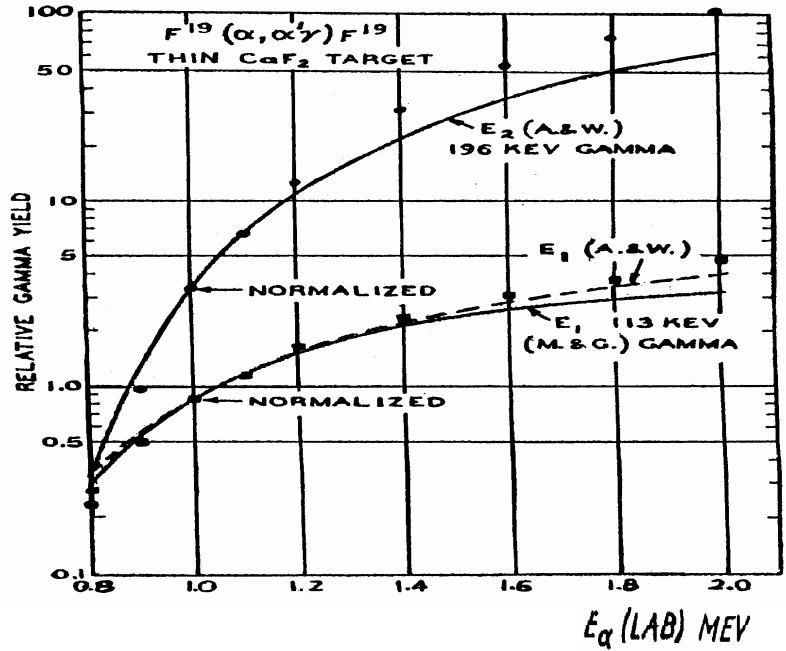

Fig. 3 : Excitation coulombienne par les particules alpha des niveaux 113 KeV et 196 KeV d'une cible mince du $^{19}$F. Les lignes pleines sont les courbes théoriques de E1 et E2 respectivement [13] et [14]. La ligne pointillée est la courbe théorique de E1 selon [13]. Les valeurs théoriques sont normalisées au point d'énergie $E_\alpha$ = 1.0 MeV

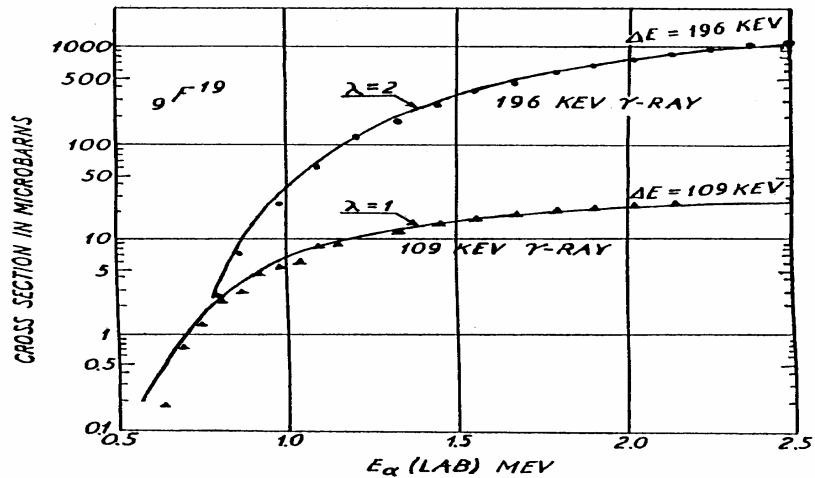

Fig. 4: Fonctions d'excitation pour les niveaux du $^{19}$F. la figure montre les sections efficaces mesurées pour E1 (109 keV) et E2 (196 keV) des rayons gamma émis par bombardement d'une cible mince CaF$_2$ [8] les fonctions théoriques sont les courbes continues, qui sont normalisées à $E_\alpha$ = 1.55 MeV

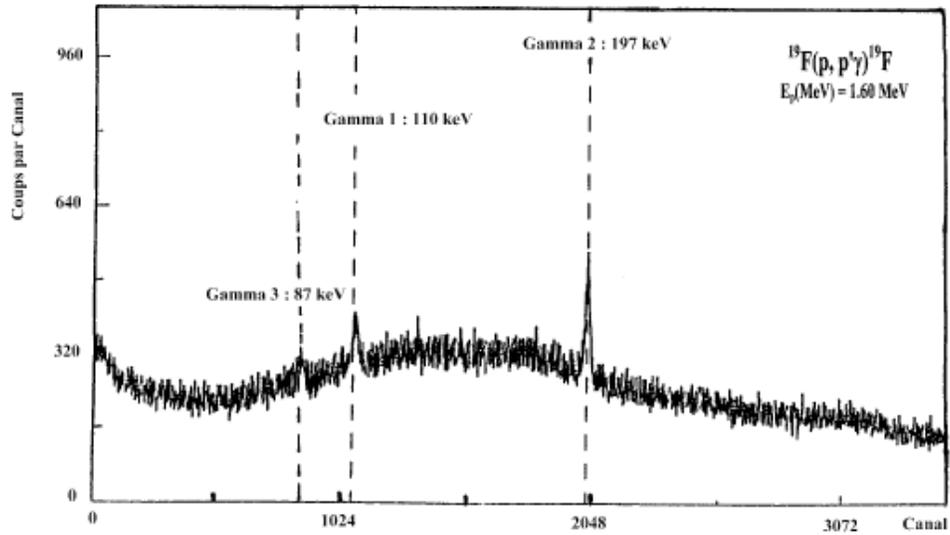

Fig. 5: Spectre des rayonnements gamma émis de l'excitation coulombienne du $^{19}$F par des protons d'énergie $E_P$ = 1.6 MeV

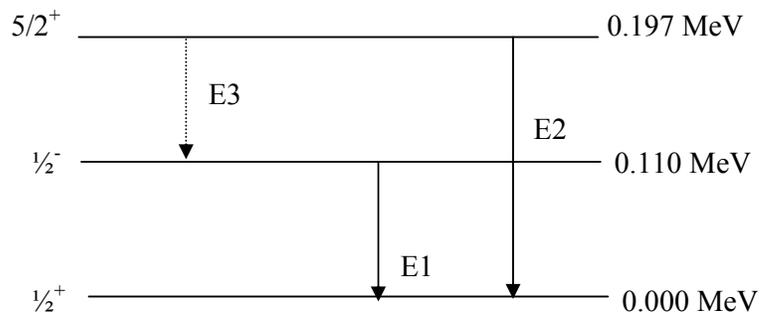

Fig. 6 : Schéma de désexcitation du $^{19}$F après l'excitation coulombienne

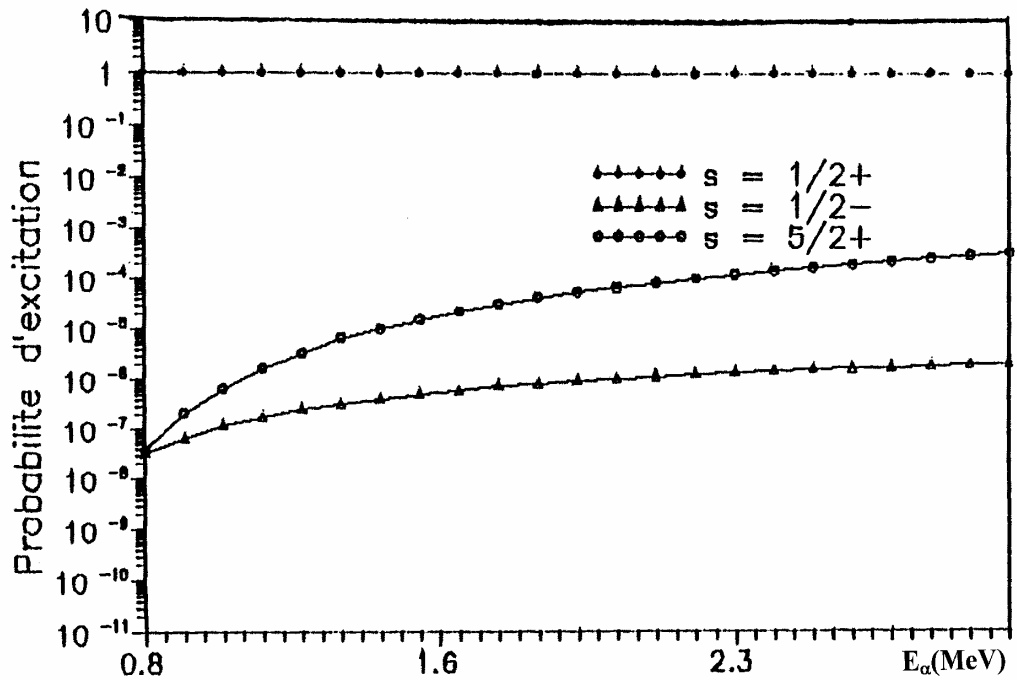

Fig. 7: Probabilités d'excitation des niveaux du $^{19}$F en fonction de l'énergie $E_\alpha$.

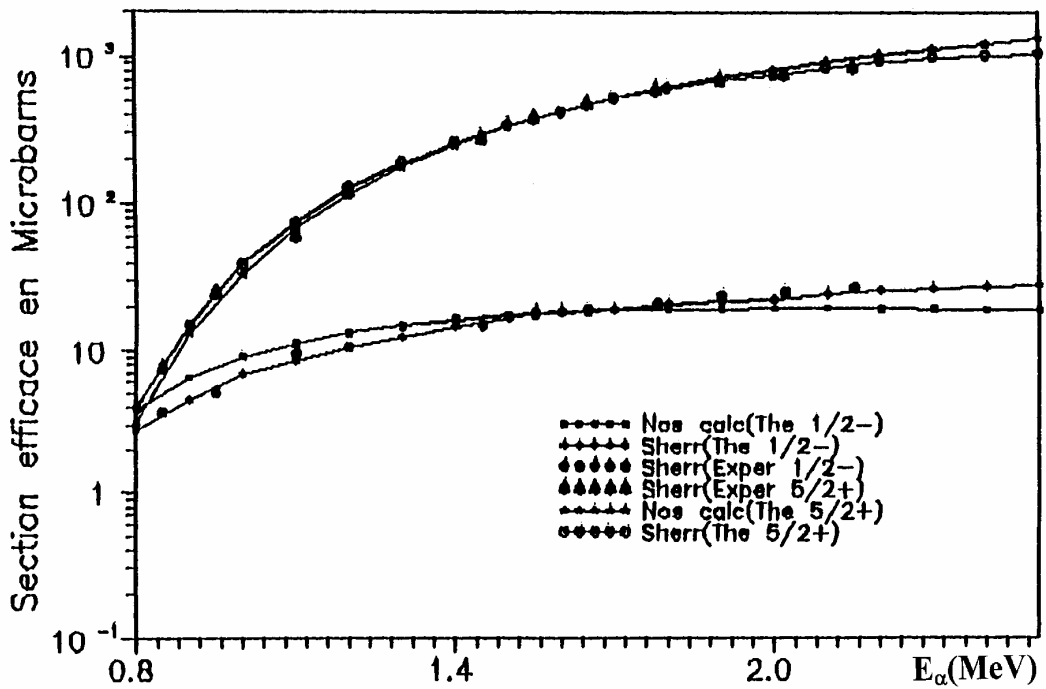

Fig. 8: Sections efficaces théoriques du $^{19}$F normalisées à $E_\alpha$ =1.55 MeV
et comparées à celles trouvées par Sherr et al [8].

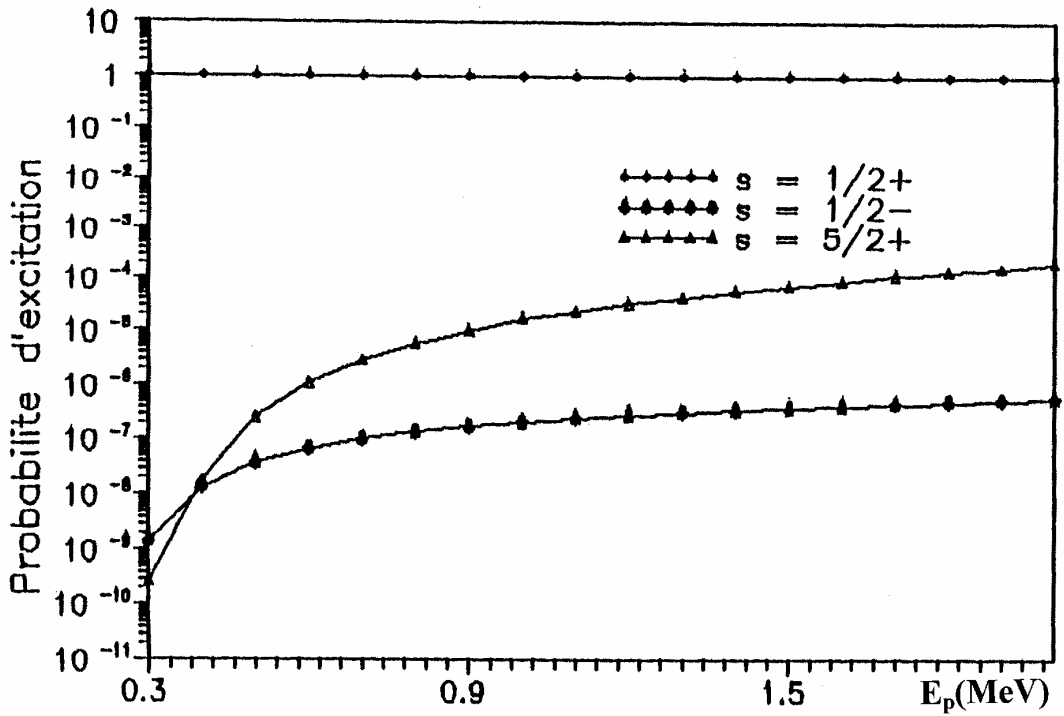

Fig. 9: Probabilités d'excitation des niveaux du $^{19}$F
en fonction de l'énergie des protons $E_P$.

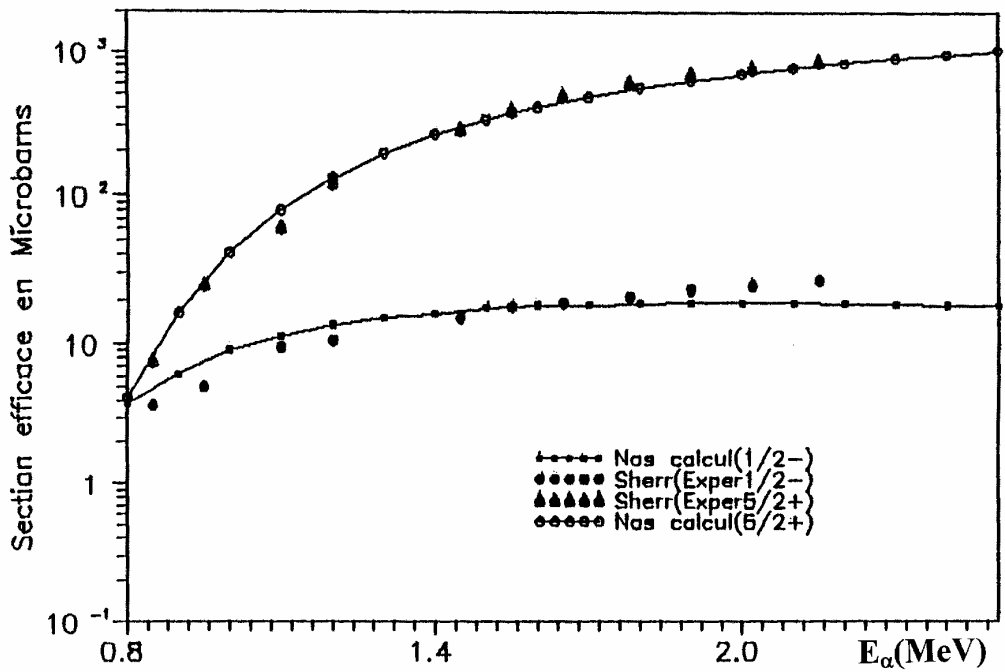

Fig. 10: Sections efficaces théoriques du $^{19}$F normalisées à $E_\alpha$ =1.55 MeV
et calculées à partir des éléments de matrice réduits que nous avons obtenu.